\documentclass[conference]{IEEEtran}
\usepackage{amsmath,amssymb,amsthm}
\usepackage{multirow}
\usepackage{graphicx}
\usepackage{stmaryrd}
\usepackage{blkarray}

\DeclareMathOperator{\F}{\mathbb F}

\DeclareMathOperator{\wt}{wt}
\theoremstyle{plain}

\newcommand{\mF}{\mathcal F}
\newcommand{\set}[1]{\left\{{#1}\right\}}
\newcommand{\ceil}[1]{\left\lceil{#1}\right\rceil}

\DeclareMathOperator{\sgn}{sgn}

\begin{document}

\title{A Randomized Construction of Polar Subcodes}
\author{
Peter Trifonov, Grigorii Trofimiuk\\
Saint Petersburg Polytechnic University\\
Email: \{petert,grigoriyt\}@dcn.icc.spbstu.ru}

\maketitle

\begin{abstract}
A method for construction of polar subcodes is presented, which aims on minimization  of the number of low-weight codewords in the obtained codes, as well as on improved performance under list or sequential decoding. Simulation results are provided, which show that the obtained codes outperform LDPC and turbo codes.
\end{abstract}

\section{Introduction}
Polar codes are a novel class of error-correcting codes, which asymptotically achieve the symmetric capacity of memoryless channels, have low complexity construction, encoding and decoding algorithms \cite{arikan2009channel}. However, the performance of polar codes of practical length is quite poor. The reasons for this are their low minimum distance and the suboptimality of the successive cancellation (SC) decoding algorithm. 

It was shown that polar codes with CRC under list decoding \cite{tal2015list} can provide the performance comparable to that of LDPC codes. Polar subcodes of extended BCH\ codes were shown to have higher minimum distance than classical polar codes and provide even better performance \cite{trifonov2016polar}.  Similar performance with much lower complexity can be obtained by employing the sequential decoding algorithm \cite{miloslavskaya2014sequential}. Experiments show that the performance of both types of codes under list/sequential decoding with small list size (e.g. $L=32$) is dominated by the events corresponding to the correct path being killed at early phases of the decoding algorithm.  

In this paper we present a randomized construction of polar subcodes, which provides substantially better performance under list/sequential decoding.  This is achieved by providing a set of dynamic freezing constraints, which allow the decoder to quickly penalize most of the incorrect paths.
The proposed construction is a heuristical one, and we do not have precise techniques for optimizing its parameters.  Nevertheless, the obtained codes outperform AR4JA LDPC codes adopted in the CCSDS standard  \cite{ccsds2011synccoding}.

\section{Background}
\label{sBackground}
\subsection{Polar codes}
A $(n=2^m,k)$ polar code over $\F_2$ is a set of vectors $c_0^{n-1}=u_0^{n-1}A_{m}$,  where $a_i^j=(a_i,\dots,a_j)$, $A_m=\begin{pmatrix}
1&0\\1&1
\end{pmatrix}^{\otimes m}B_m$ is a matrix of the polarizing transformation, $B_m$ is the bit-reversal permutation matrix,    $F^{\otimes m}$ denotes $m$-times Kronecker product of matrix $F$ with itself, $u_i=0, i\in \mF$,  $\mF\subset[n]$ is a set of $n-k$ frozen symbol indices, and $[n]=\set{0,\dots,n-1}$.
It can be seen that the minimum distance of a polar code is given by $2^r$, where $r=\min_{i\in [n]\setminus \mF} \wt(i)$. 


It is possible to show that  matrix $A_m$ together with a memoryless output symmetric channel $\mathbf W(y|x)$ gives rise to synthetic bit subchannels with transition probability functions $$\mathbf W_m^{(i)}(y_0^{n-1},u_0^{i-1}|u_i)=\frac{1}{2^{n-1}}\sum_{u_{i+1}^{n-1}}\prod_{j=0}^{n-1}\mathbf W(y_j|(u_0^{n-1}A_{m})_j).$$
One can compute the capacity $C_{m,i}$ and bit error rate $P_{m,i}$ in each of the subchannels $\mathbf W_m^{(i)}$ using the techniques presented in \cite{tal2011how,trifonov2012efficient}. Classical polar codes are obtained by taking $\mF$ to be the set of $n-k$ indices $i$ of bit subchannels $\mathbf W_m^{(i)}$ with low capacity or high error probability. 
\subsection{Polar subcodes}

It was suggested in \cite{trifonov2016polar} to set frozen symbols not to zero, but to some linear combinations of other symbols, i.e. 
\begin{equation}
\label{mDynFrozen}
u_i=\sum_{j<i}V_{s_i,j}u_j,i\in \mF,
\end{equation}
where $V$ is a $(n-k)\times n$ constraint matrix, such that distinct rows end\footnote{Given some binary vector $a_0^{n-1}$, we say that it  ends in position $j$ iff $a_j=1$ and $a_t=0,  j<t<n$.
 } in distinct columns $i\in \mF$, and $s_i$ is the index of the row ending in column $i$.  Symbols $u_i$ with at least one term in the r.h.s. of \eqref{mDynFrozen} are referred to as dynamic frozen (DFS), and those with $V_{s_i,j}=0,j<s_i,$ are denoted static frozen. Matrix $V$ can be constructed so that codewords $c_0^{n-1}$ belong to some $(n,k'>k,d)$ {\em parent code} (e.g. extended BCH) with sufficiently high minimum distance $d$, and the SC decoding error probability $P(\mF)$ is minimized. The obtained codes are referred to as polar subcodes. 

The successive cancellation (SC)\ decoding algorithm for polar subcodes makes decisions 
\begin{equation}
\label{mSC}
\hat u_i=\begin{cases}
\arg \max_{u_i\in \F_2}\mathbf W_m^{(i)}(y_0^{n-1},\hat u_0^{i-1}|u_i),&i\notin \mF\\
\sum_{j<i}V_{s_i,j}\hat u_j,&i\in \mF.
\end{cases}
\end{equation}

\subsection{List decoding}
The original SC decoding algorithm does not provide maximum likelihood (ML) decoding of polar codes. The Tal-Vardy list decoding algorithm \cite{tal2015list} was shown to provide substantially better performance. In this section we review the min-sum version of this algorithm \cite{balatsoukasstimming2015llrbased}.

Let $y_0^{n-1}$ be the channel output vector.
The decoder keeps $L$ partial information vectors $\hat u_0^{i-1}$ together with their scores $R(\hat u_0^{i-1},y_0^{n-1})$. At  phase $i, 0\leq i<n,$ the decoder constructs at most $2L$ possible continuations $\hat u_0^i$ of these vectors, and computes their scores. Then $L$ vectors $\hat u_0^i$ with the highest scores are selected, and the remaining ones are killed. At phase $n$ the algorithm returns vector $\hat u_0^{n-1}$ with the highest score.
 
The score of vector $\hat u_0^i$ is given by 
\begin{equation}
\label{mScore}
R(\hat u_0^{i},y_0^{n-1})=R(\hat u_0^{i-1},y_0^{n-1})+\tau(\hat u_i,S_m^{(i)}(\hat u_0^{i-1},y_0^{n-1})),
\end{equation} 
where $$\tau(u,S)=\begin{cases}
0,&\text{if $(-1)^u=\sgn S$}\\
-|S|,&\text{otherwise}
\end{cases}
$$
is the penalty function, 
and $S_m^{(i)}(\hat u_0^{i-1},y_0^{n-1})$ are the approximate log-likelihood ratios, which are given by
\begin{align}
\label{mMinSum1}
S_{\lambda}^{(2i)}(\hat u_0^{2i-1},y_0^{N-1})=&\sgn (a)\sgn (b)\min(|a|,|b|),\\
\label{mMinSum2}
S_{\lambda}^{(2i+1)}(\hat u_0^{2i},y_0^{N-1})=&(-1)^{\hat u_{2i}}a+b,
\end{align}
where $a=S_{\lambda-1}^{(i)}(\hat u_{0,e}^{2i-1}\oplus \hat u_{0,o}^{2i-1},y_0^{\frac{N}{2}-1})$, $b=S_{\lambda-1}^{(i)}(\hat u_{0,o}^{2i-1},y_{\frac{N}{2}}^{N-1})$, and $S_0^{(0)}(y)=\log\frac{\mathbf W(y|0)}{\mathbf W(y|1)}$. 

The complexity of this algorithm can be substantially reduced by employing sequential decoding techniques \cite{miloslavskaya2014sequential}.

The error probability of the Tal-Vardy algorithm is given by $$P(L)=P_{ML}+P(E(L)|\mathbb C),$$
where $P_{ML}$ is the maximum likelihood decoding error probability, $\mathbb C$ is the event corresponding to the maximum likelihood decoder producing the correct codeword $u_0^{n-1}A_m$, and $E(L)$ is the event corresponding to the score $R(u_0^i,y_0^{n-1})$ of the correct vector becoming lower than the scores of $L$ incorrect vectors $\hat u_0^i$ at some intermediate phase, so that the correct vector is killed by the decoder. The value of $P_{ML}$ can be estimated using  the weight distribution of the code. At high SNR $P_{ML}$  primarily depends on the minimum distance $d$ and error coefficient of the code, i.e. the number $w_d$ of codewords of weight $d$. To the best of our knowledge, there are still no techniques for computing $P(E(L)|\mathbb C)$. However, experiments show that this quantity increases with $P_{SC}$, the successive cancellation decoding error probability of the code.

\section{A randomized code construction}
\label{sConstruction}
Polar subcodes of extended BCH codes were shown to provide substantially better performance compared to  classical polar codes under list and sequential SC decoding \cite{trifonov2016polar}. It turns out that list/sequential decoding with very large list size $L$ is needed in order to implement near-ML\ decoding of polar subcodes with even modestly high minimum distance $d$. For small $L$ almost all error events are caused by the decoder killing the correct vector $u_0^i$ at early phases. Therefore, we propose a code construction, which is   targeted to be efficiently decodable by list/sequential algorithms with small $L$.

\subsection{Eliminating low-weight codewords from a polar code}
In order to obtain a $(n,k)$ polar subcode $\mathcal C$, we propose to construct a $(n,k+t)$ classical Arikan polar code $\mathcal C'$, called {\em base code}, and then select a random $k$-dimension linear subspace of this code.  Let $d'$ be the minimum distance of $\mathcal C'$, and let $w_i'$ be its weight distribution.  
It is possible to show \cite{Zubkov2014Probabilistic} that the expected number of codewords of weight $s$ in $\mathcal C$ is given by  
\begin{equation}
\label{mExpSpectrum}
\mathbf E[w_s]=w_s'\frac{2^k-1}{2^{k+t}-1}\approx w_s'2^{-t},s>0,
\end{equation}
provided that all  linear subspaces are selected equiprobably.

The parameter $t$ should be selected  so that the expected number $\mathbf E[w_{d'}]$ of codewords of weight $d'$ in code $\mathcal C$ becomes  sufficiently small. If $\mathbf E[w_{d'}]<1$, then with high probability  the minimum distance $d$ of $\mathcal C$ is higher than $d'$.

It was shown in \cite{bardet2016algebraic} that the error coefficient of a classical polar code $\mathcal C'$ of length $n=2^m$ with the set of frozen symbols $\mF'$, i.e. the number of codewords of weight $d'$, is given by 

\begin{equation}
\label{mPolarErrCoeff}
w_{d'}'=2^{m-r}\sum_{\substack{g\in [n]\setminus \mF'\\\wt(g)=r}}2^{|\lambda_g|},
\end{equation}
where $r=\min_{i\in [n]\setminus \mF'}\wt(i)$, $\lambda_g$ is the list of indices of zero bits in integer $g$, and $|(i_{0},\dots,i_{m-r-1})|=\sum_{j=0}^{m-r-1}(i_j-j)$.
It can be also seen that any  codeword of $\mathcal C'$ of weight $d'=2^r$ is obtained as a linear combination of a weight-$d'$ row of matrix $A_m$, and possibly some other rows of this matrix.

Taking a random linear subcode of $\mathcal C'$ is equivalent to selecting codewords of $c=uA_m\in \mathcal C'$, which satisfy $c\tilde H^T=0$, where $\tilde H$ is a $t\times n$ matrix randomly selected from the set of full-rank binary matrices.  Alternatively, one can randomly select a full-rank $t\times n$ matrix  $\tilde V=\tilde H A_m^T$, so that the constraint matrix of code $\mathcal C$ is given by $V=\begin{pmatrix}V'\\\tilde V\end{pmatrix}$, where $V'$ is the constraint matrix of code $\mathcal C'$ consisting of weight-1 rows. It can be assumed without loss of generality that the positions of last non-zero elements in the rows of $V$ are distinct.

In order to simplify the implementation, as well as to obtain codes better decodable by the Tal-Vardy list decoding algorithm, we propose to impose dynamic freezing constraints \eqref{mDynFrozen} onto symbols $u_i$, such that $i$ are the smallest possible values, and all non-frozen symbols $u_i: \wt(i)=r$, participate in \eqref{mDynFrozen} with high probability. Since any weight-$2^r$ codeword $c_0^{n-1}=u_0^{n-1}A_m$ corresponds to $u_0^{n-1}$ having $u_i=1$ for some $i:\wt(i)=r$, the latter requirement ensures that most of the low-weight codewords are eliminated from $\mathcal C'$. The former requirement enables the list decoder to process the corresponding constraints at the earliest possible phases, reducing thus the probability of the correct vector  being killed. Employing this approach, however, causes \eqref{mExpSpectrum} to be an imprecise estimate of the weight distribution components of the obtained code.

The constraints obtained with this approach are referred to as type-A dynamic freezing constraints (DFC-A), and the corresponding symbols in l.h.s of \eqref{mDynFrozen} are called type-A DFS. 
\subsection{Design of codes with better list decodability}
Let $u_0^{n-1}$ be the vector corresponding to the transmitted codeword. Consider the case of $\tau(u_j,S_m^{(j)}(u_0^{j-1},y_0^{n-1}))<0$ for some $j\notin\mF$, i.e. an error event of the SC algorithm. Such event causes the Tal-Vardy list decoder to consider an incorrect vector $\hat u_0^{j}$ with $R(\hat  u_0^{j},y_0^{n-1})>R(u_0^{j},y_0^{n-1})$.  

It was observed in \cite{arikan2011systematic} that classical polar codes exhibit very low bit error rate. Indeed, most of erroneous received symbols correspond to low-magnitude log-likelihood ratios. Hence, with sufficiently high probability the sign of the LLRs $S_m^{(i)}(\hat u_0^{i-1},y_0^{n-1}), i>j,$ obtained from \eqref{mMinSum2} may be correct (i.e. agree with the value of $u_i$) even if $\hat u_j\neq u_j$ for some $j<i$. Hence, the decoder may need to process many frozen symbols before the score of an incorrect path $\hat u_0^{i}$ is sufficiently penalized according to \eqref{mScore}.  It may happen that the score $R(u_0^i,y_0^{n-1})$ becomes lower than the scores of $L$ incorrect paths, so that the Tal-Vardy list decoder kills $u_0^i$, i.e. makes an error.

In order to reduce probability of such events we propose to select $q$ symbols $u_i$ corresponding to most reliable bit subchannels, which would be static frozen in the classical polar code construction method, and impose on them dynamic freezing constraints with random coefficients. Such constraints are referred to as type-B dynamic freezing constraints (DFC-B), and the corresponding $u_i$ are denoted type-B dynamic frozen symbols (DFS-B).

Let $u_i$ be a  DFS-B. For an incorrect path $\hat u_0^i$ with high probability one obtains $u_i\neq \sum_{j<i}V_{s_i,j}\hat u_j$, so that the r.h.s. values of this expression does not agree with the sign of $S_m^{(i)}(\hat u_0^{i-1},y_0^{n-1})$. This causes the scores  $R(\hat u_0^i,y_0^{n-1})$ of most incorrect paths $\hat u_0^{i}$ to quickly decrease with $i$.

\subsection{Shortening}
The classical construction of polar codes is limited to length $2^m$. However, practical systems require codes of other lengths $n$. Shortening can be used to obtain codes of arbitrary length. For the sake of simplicity, we do not consider optimization of a shortening pattern, and use the method suggested in \cite{wang2014punctured}. Namely, we set $u_i=0, n\leq i<2^m$ (shortening constraints), and eliminate from the vector $u_0^{2^m-1}A_m$ symbols with indices $r_m(i)$, where $$r_m\left(\sum_{j=0}^{m-1}i_j2^j\right)=\sum_{j=0}^{m-1}i_{m-1-j}2^j$$
is the bit reversal function. By examining the matrix $A_m$, one can see that these symbols are equal to zero. Observe that shortening should be taken into account while evaluating the reliability of  bit subchannels $\mathbf W_m^{(i)}$.

\subsection{The proposed code construction algorithm}

Below we present a summary of the proposed method for construction of an $(n,k)$ randomized polar subcode. Here $t$ and $q$ denote the number of type-A and type-B dynamic freezing constraints, respectively.

\begin{enumerate}
\item        Let $V$ be a $(2^m-k)\times 2^m$ matrix initially filled with 0. 
\item        \,[{\bf Shortening constraints}] Let $s=2^m-n$. Set $V_{0\dots s-1,n\dots 2^m-1}$ to a $s\times s$ identity matrix.
\item        Let  $\mF\subset [n]$ be the set of  $n-k-t$ indices of least reliable bit subchannels  $\mathbf W_m^{(j)}, 0\leq j<n,$ induced by $A_m$. Let $\mathcal N=\left[n\right]\setminus \mF$. Let  $\hat{\mathcal F}$ be the set of $q$ elements of $\mF$ corresponding to the most reliable subchannels, and let $\tilde {\mathcal F}=\mF\setminus \hat{\mathcal F}$. 
\item        Let $w=\min_{i\in\mathcal N}\wt(i)$ be the minimal weight of indices of type-A dynamic frozen symbols.
\item        Let $Z=\set{z_0,\dots,z_{t-1} }\subset \mathcal N$, be the set of $t$ integers, which is constructed by selecting from $\mathcal N$ maximal integers  of weight $w, w+1,$ etc, until $t$ integers are selected. 
\item        \,[{\bf DFC-A}] For $j$ from  $0$ to $t-1$ do:
\begin{enumerate}
\item                 Set $V_{s,z_j }:=1$. 
\item Set $V_{s,i},i\in \mathcal N,i<z_j$ to independent random equiprobable binary values (IREPBV). 
\item Let $s:=s+1$.
        \end{enumerate}
\item      \,[{\bf Static freezing constraints}]   For $i$ from  $0$ to $n-k-q-t-1$ do
\begin{enumerate}
\item        Set $V_{s,f_i }:=1$,  where $f_i$ is the $i$-th element of $\tilde F$. 
\item Let $s:=s+1$. 
        \end{enumerate}
\item        \,[{\bf DFC-B}] For $i$ from  $0$ to $q-1$ do
\begin{enumerate}
\item                Let $V_{s,f_i }:=1$,  where $f_i$ is the $i$-th element of $\hat F $.
\item        Set $V_{s,j},j\in\mathcal N,j<f_i$ to IREPBV.
\item        Let $s:=s+1$.
        \end{enumerate}
\end{enumerate}
In practice, the elements of $V_{s,j}$ at steps 6.b and 8.b in the above algorithm can be obtained from a pseudo-random number generator (PRNG). Hence, the code can be completely specified by the seed value of the PRNG,  parameters of the channel $\mathbf W(y|x)$, and numbers $n,k,t,q$.
 It can be seen that the complexity of this algorithm is $O(k(t+q))$ calls to the PRNG.
 Recall, that construction of a constraint matrix of a polar subcode of an $(n,\tilde k,d)$  extended BCH\ code with $(n-\tilde k)\times $ check matrix $\tilde H$ requires Gaussian elimination to be performed on matrix $\tilde HA_m^T$, which requires $(n-\tilde k)^2n$ operations \cite{trifonov2016polar}.  

Due to lack of analytic methods for performance evaluation of polar codes under list/sequential decoding, we are not able to give  optimal values of  $t$  and $q$. However, experiments suggest that setting $t=\min(m,n-k)$, $m=\ceil {\log_2 n}$ and 
\begin{equation}
\label{mDFSB}
q=\max(0,\min(Q-t,n-k-t)),
\end{equation}
 where $Q=64$, results in codes with good performance under list decoding with $L\approx 32$. In some cases better codes can be obtained by careful selection of these parameters.

\section{Numeric results}
\label{sNumRes}
\begin{table}
\caption{Error coefficient of $(1024,512)$ randomized polar subcodes}
\label{tErrCoeff1024512}
\scalebox{0.82}{
\parbox{0.8\textwidth}{
\begin{tabular}{|c|c|c|c|c|c|c|c|c|c|}\hline
\multirow{2}{*}{$t$}&\multirow{2}{*}{$w_{16}'$}&\multirow{2}{*}{$P_{SC}$}&\multirow{2}{*}{$\mathbf E[w_{16}]$}&\multicolumn{3}{c|}{$q=0$}&\multicolumn{3}{c|}{$q=64-t$}\\\cline{5-10}
&&&&$\min w_{16}$&$\max w_{16}$&$\overline w_{16}$&$\min w_{16}$&$\max w_{16}$&$\overline w_{16}$\\\hline
1&53440&0.231&26720&27064&27360&27214.9&20056&21472&20866.4\\\hline
2&54464&0.234&13616&13416&13896&13560.9&9872&10932&10381.5\\\hline
6&54464&0.261&851&910&1072&988.04&568&753&661.68\\\hline
9&66752&0.270&130.4&183&283&226.78&111&184&145.18\\\hline
10&66752&0.277&65.2& 84&148&109.08&46&112&73.72\\\hline
11&66752&0.290&32.6&15&67&41.1&14&57&35.3\\\hline
16&91328&0.314&1.39&0&10&3.76&0&10&2.98\\\hline
\end{tabular}}}

\end{table}

Table \ref{tErrCoeff1024512} illustrates the error coefficient of  $(n=1024,k=512)$ randomized polar subcodes constructed for AWGN\ channel with BPSK modulation and $E_b/N_0=1.5 $ dB.   For each $t$ and each $q$ 50 codes were constructed,  and their low-weight codewords  were obtained using the algorithm given in \cite{canteaut98new}. In all cases the minimum distance of the constructed codes was at least $16$.   We report the error coefficient $w_{16}'$ of the corresponding $(n,k+t)$ base code given by \eqref{mPolarErrCoeff}, the successive cancellation decoding error probability $P_{SC,}$ and the minimal, maximal and average values of $w_{16}$ for the constructed codes, denoted  $\min w_{16},\max w_{16},\overline w_{16}$, respectively.  It can be seen that for small $t$ and $q=0$ the value of $\overline w_{16}$ is quite close to that predicted by \eqref{mExpSpectrum}.
However, for $t>7$ and $q=0$ the average error coefficient of the constructed codes is considerably higher than $\mathbf E[w_{16}]$ for a uniformly selected subcode of the base code. The reason for this is that the above presented code construction algorithm does not implement equiprobable selection of $k$-dimensional linear subspaces from the $(k+t)$-dimensional classical polar code $\mathcal C'$. It can be seen that the codes obtained with $q>0$ (which are not subcodes of $\mathcal C'$) have substantially lower $\overline w_{16}$.   

In both cases the average error coefficient $\overline w_{16}$ quickly decreases with $t$. However, increasing $t$ requires one to unfreeze some symbols corresponding to unreliable bit subchannels, so $P_{SC}$ and $P(E(L)|\mathbb C)$ increase with $t$, and at some point the performance becomes dominated by the suboptimality of the list/sequential decoding algorithm.
\begin{figure}
\includegraphics[width=0.5\textwidth]{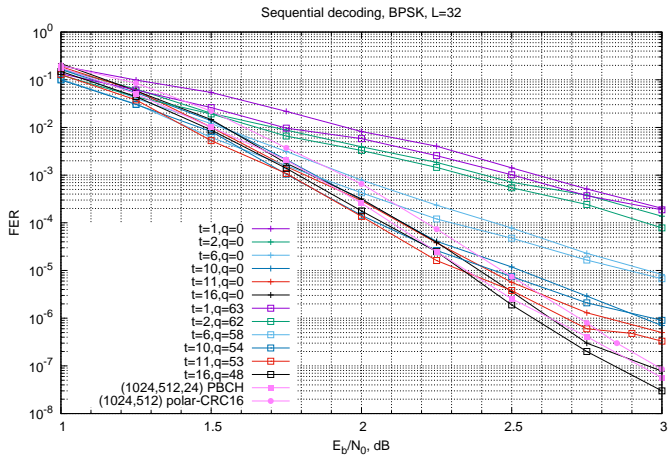}
\caption{Performance of $(1024,512)$ randomized   polar    subcodes}
\label{f1024512}
\end{figure}
Figure \ref{f1024512} illustrates the performance of the considered codes in the case of AWGN channel with BPSK\ modulation and sequential decoding \cite{miloslavskaya2014sequential}. For each combination of $t,q$ a code was selected with $w_{16}\approx\overline w_{16}$. For comparison, we report also the performance of the $(1024,512,24)$ polar subcode of an extended BCH code (PBCH) and a polar code with CRC \cite{trifonov2016polar}, It can be seen that the codes with $q>0$ outperform the corresponding codes with $q=0$, and 
the best performance in the low-SNR region is provided by the code with $t=11,q=53$. For $E_b/N_0>2$ dB the best performance is obtained in the case of $t=16,q=48$ due to much lower error coefficient of the corresponding code.
For $E_b/N_0 \leq 2.1$ dB the randomized polar subcode with $t=11,q=53$ provides up to 0.1 dB gain  with respect to the PBCH\ code and 0.2 dB gain compared to the polar code with CRC-16. Observe, that the code obtained with $t=16,q=0$ also outperforms the polar code with CRC. The latter code can be considered as an instance of the proposed construction with $q=0$ and $Z=\set{n-c,\dots,n-1}$, where $c=16$ is the number of bits in CRC. The reason for this gain is that the proposed approach enables the decoder to process the dynamic freezing constraints at earlier phases, reducing thus the probability of the correct vector being killed.

\begin{figure}[t]
\includegraphics[width=0.5\textwidth]{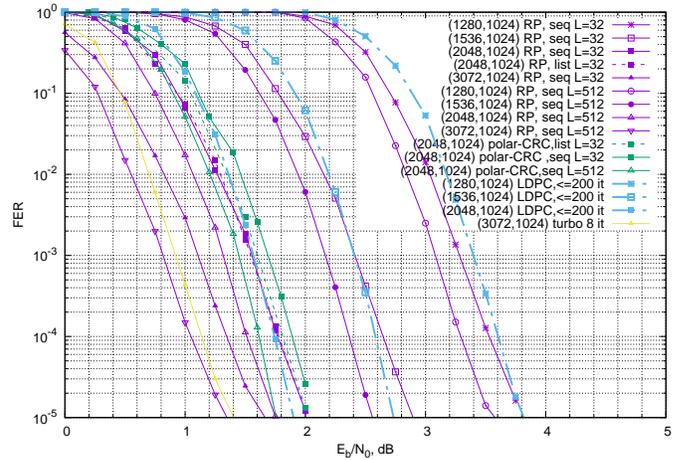}
\caption{Performance  of codes    with $k=1024 $}
\label{fPerf1024}
\end{figure}
\begin{figure}
\includegraphics[width=0.5\textwidth]{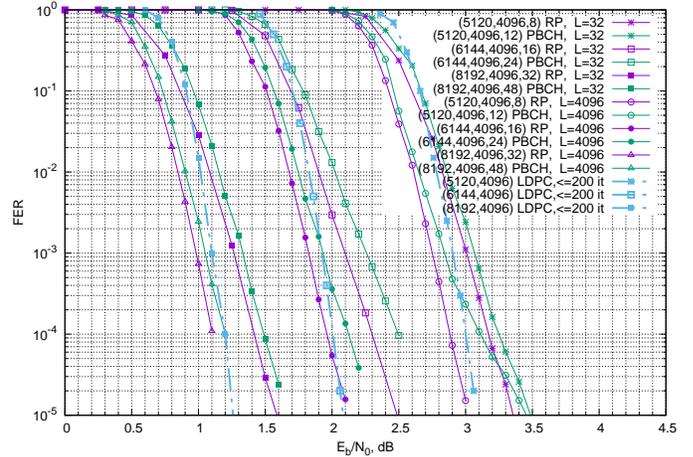}
\caption{Performance of codes with $k=4096 $}
\label{fPerf4096}
\end{figure}
\begin{figure}
\includegraphics[width=0.5\textwidth]{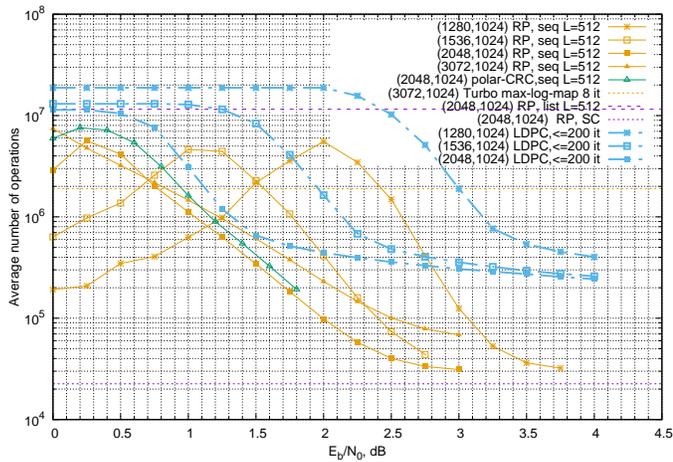}
\caption{Decoding complexity of codes  with  $k=1024 $}
\label{fPerf1024Compl}
\end{figure}
Figures \ref{fPerf1024} and \ref{fPerf4096} illustrate the performance of  polar subcodes of dimension $k=1024$ and $k=4096$.  We report the results for the case of the proposed randomized polar subcodes (RP), polar subcodes of eBCH\ codes (PBCH), and a polar code with  CRC-16.  For comparison, we provide also the results for AR4JA LDPC codes \cite{ccsds2011synccoding} under shuffled BP decoding \cite{zhang2005shuffled} and an LTE turbo code. It can be seen that for $k=1024$ and $L=32$ the proposed randomized polar subcodes outperform the polar code with CRC, and provide performance comparable to that of LDPC codes. For $L=512$ the proposed codes provide up to 0.5 dB gain with respect to LDPC, turbo and polar-CRC codes.  In the case of $k=4096$  list size $L$ needs to be increased in order to obtain gain with respect to the LDPC codes.  The reason is that  although the fraction of mediocre (i.e. those having Bhattacharyya parameters $0<a<Z_m^{(i)}<b<1$ for some fixed $a,b$)  bit subchannels decreases with $m$, the absolute number of such subchannels corresponding to unfrozen symbols, which with high probability cause $E(L)$ events, increases with $m$ \cite{hassani2014finitelength}. Observe that the $(2048,1024)$ polar code with CRC exhibits noticeable performance loss in the case of sequential decoding. However, the performance of the proposed randomized polar subcodes is essentially the same in the case of both decoding algorithms. 

Figure \ref{fPerf1024Compl} illustrates the average number of operations performed by the decoders  of the considered codes. For polar and turbo codes, decoding involves only summations and comparisons, while for LDPC codes the algorithm in \cite{zhang2005shuffled} makes use of  summations and $\log\tanh(\cdot)$ function. It can be seen that the $(2048,1024)$ randomized polar subcode provides not only better performance, but also lower decoding complexity compared to the polar code with CRC. A surprising result is that randomized polar subcodes outperform polar subcodes of eBCH\ codes with much higher minimum distance even in the high-SNR region.
\begin{figure}
\centering
\includegraphics[width=0.5\textwidth]{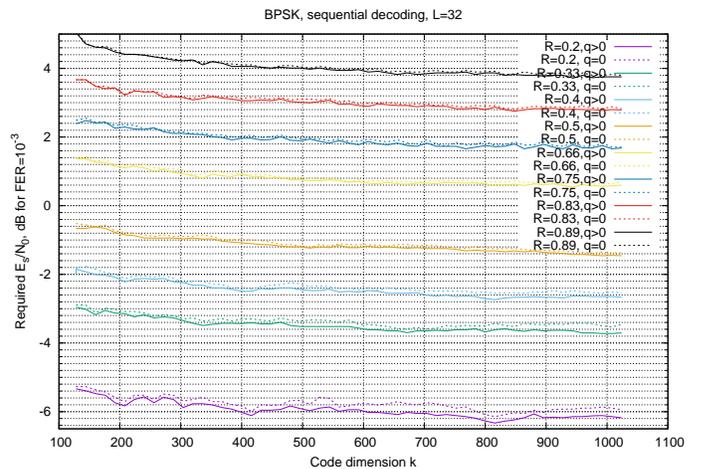}
\caption{The gain of type-B dynamic frozen symbols }
\label{fLDPCFight}
\end{figure}

Figure \ref{fLDPCFight} illustrates the SNR needed for achieving codeword error rate $10^{-3}$  in the case of BPSK\ modulation for different values of code rate $R=k/n$ and dimension  $k$ for the case of $q=0$ and $q$ given by  \eqref{mDFSB}. It can be seen that employing DFS-B provides up to 0.3 dB gain, especially for low-rate codes.

\section{Conclusions}
In this paper a  randomized construction of polar subcodes was proposed. The construction relies on two types of dynamic freezing constraints, which allow one to reduce the number of low-weight codewords in the obtained code and decrease the probability of a correct path being killed by the Tal-Vardy list decoder. The obtained codes were shown to outperform LDPC and turbo codes, and have  lower complexity in the case of sequential decoding compared to polar codes with CRC. 

The proposed construction is a heuristical one. It relies on some observations on the behaviour of the Tal-Vardy list decoding algorithm. Any progress in its analytic performance evaluation may result in an improved code design.

\bibliographystyle{ieeetran}

\end{document}